\documentclass[prl,showkeys,twocolumn]{revtex4}

\usepackage{graphicx,amssymb,amsmath}
\usepackage[colorlinks,bookmarksopen,bookmarksnumbered,citecolor=red,urlcolor=red]{hyperref}

\newcommand\BibTeX{{\rmfamily B\kern-.05em \textsc{i\kern-.025em b}\kern-.08em
T\kern-.1667em\lower.7ex\hbox{E}\kern-.125emX}}

\begin{document}


\title{The spatial organization of the population density in cities}

\author{Valerio Volpati}
\affiliation{Institut de Physique Th\'{e}orique, CEA, CNRS-URA 2306, F-91191,
Gif-sur-Yvette, France}
\author{Marc Barthelemy}
\email{marc.barthelemy@ipht.fr}
\affiliation{Institut de Physique Th\'{e}orique, CEA, CNRS-URA 2306, F-91191,
Gif-sur-Yvette, France}
\affiliation{Centre d'Analyse et de Math\'ematique Sociales, (CNRS/EHESS) 54 Boulevard Raspail, 75006 Paris, France}


\begin{abstract}

Although the average population density of a city is an extremely simple indicator, it is often used as a determinant factor for describing various aspects of urban phenomena. On the other hand, a plethora of different measures that aim at characterizing the urban form have been introduced in the literature, often with the risk of redundancy. Here, we argue that two measures are enough to capture a wealth of different forms of the population density.  First, fluctuations of the local density can be very important and we should distinguish almost homogeneous cities from highly heterogeneous ones. This is easily characterized by an indicator such as the Gini coefficient $G$, or equivalently by the relative standard deviation or the entropy. The second important dimension is the spatial organization of the heterogeneities in population density. We propose a dispersion index $\eta$ that characterizes the degree of localization of highly populated areas. As far as population density is concerned, we argue that these two dimensions are enough to characterize the spatial organization of cities.  We discuss this approach using a dataset of about $4,500$ cities belonging to the $10$ largest urban areas in France, for which we have high resolution data, at the level of a square grid of $200\times 200$ meters. Representing cities in the plane $(G,\eta)$ allows us to construct families of cities. We find that, on average, compactness increases with heterogeneity. More precisely, we find four large categories of cities (with population larger than $10,000$ inhabitants): (i) first, homogeneous and dispersed cities where the density fluctuations are small, (ii) very heterogeneous cities with a compact organization of large densities areas. The last two groups comprise heterogeneous cities with (iii) a monocentric organization or (iv) a more delocalized, polycentric structure. We believe that integrating these two parameters in econometric analysis could improve our understanding of the impact of urban form on various socio-economical aspects.

\end{abstract}


\maketitle

\section{Introduction}

Understanding what are the factors that affects the sustainability, the resilience or other crucial aspects of cities is 
a major goal in scientific approaches to urbanism \cite{Batty:2013,Barthelemy:2016}. In order to make decisions for improving cities, urban planners and policy makers rely on few indicators that are determinant for many socio-economical processes (see for example \cite{Li:2018} for a discussion about urban form and productivity, \cite{Fang:2015} about urban form and CO2 emissions). Obviously in the potential list of major determinants, we find the infrastructure, the spatial organization of activities and buildings, and in particular, the distribution of the local population density. At a very coarse-grained level, it has been argued some time ago that the gasoline consumption in different cities, is a simple decreasing function of average urban density \cite{Newman:1989}. Even if this result has been challenged in recent studies\cite{Louf:2013,Louf:2014}, it triggered a large amount of discussions about the need to advocate for cities to be more compact and to stop urban sprawling \cite{Ewing:2015,Batty:2003}. This example illustrates the importance of being able to characterize the urban form, and the possible implication for socio-economical processes. 

Characterizing urban form is a difficult task. Very generally, reducing the information contained in a spatial distribution to s few numbers is a complex problem without a clear solution. In the literature we can find a wealth of studies proposing various indices, built with the aim of quantifying the compactness or spreadness of population density \cite{Ewing:2015} or to characterize the monocentricity or polycentricity of the urban form \cite{Bertaud:2004}. While the first stylized theoretical studie in Spatial Economics were predicting monocentric structure for activities in cities \cite{vonThunen:1826,Alonso:1964,Fujita:1999}, modern cities often quite do not fit in this picture. It has been reported that the impact of housing prices and congestion has been driving cities towards a transition between monocentric and polycentric structures \cite{Bertaud:2004, Louf:2014}. For instance, the monocentricity of cities has been quantified by measuring population-density gradients from the center of the urban areas \cite{Bertaud:2003,Guerois:2008}. Typically, according to these measures, the monocentric description fits quite well the population distribution in several urban areas, though different functions have been proposed to describe how population density decreases, when moving away from the center \cite{Guerois:2008}. In \cite{Bertaud:2003}, these measures are used to discuss different reasons why the monocentricty picture often fails. In these studies however, typically urban (or metropolitan) areas are considered and not individual municipalites. 

The problem of `urban sprawl' is somehow similar to the monocentric polycentric urban form, and many attempts to quantify the compactedness or spreadness of cities have been proposed. In particular, Tsai \cite{Tsai:2005} developed a set of variables to distinguish compactness from `sprawl', and identified four dimensions that have a special importance for characterizing cities: (i) the total population, (ii) the average population density, (iii) the heterogeneities in population density, and (iv) the spatial structure of these heterogeneities. In this study, the Moran coefficient (a measure of spatial correlation \cite{Moran:1950}) is used in order to characterize the spatial structure of population distribution. However, such an indicator can be quite problematic: as also pointed out in Tsai \cite{Tsai:2005}, a low Moran value might imply either a high level of sprawl, or a discontinuous development. Very different urban form might then have similar values for this indicator. This is due to the definition of the Moran, which is a local quantity that depends essentially on how many highly populated areas are neighbours of low populated areas. indeed, many studies using the Moran coefficient discuss very local aspects of cities, such as noise exposures of building \cite{Silva:2013}, or the smog production of cities \cite{Liu:2017}, which depends on the micro-structure of cities (more precisely, how building and roads are distributed). Similarly, in a recent study \cite{Sobstyl:2018}, the decay of the correlation functions for the density of buildings with distance, is used to quantify the impact of urban form on temperature in cities. In the same spirit of the Moran coefficient, similar measures were proposed in order to quantify the spread in space of some activities. In \cite{Marcon:2009} for example, the so-called M-functions are introduced in order to measure the spatial concentration of firms in a city.

Another aspect of \cite{Tsai:2005}, is that all these various parameters are a priori not independent and a crucial goal would be to reduce their number and to eliminate redundancies. In particular, it is important not to specialize to specific organizations and to be able to characterize very generally cities and to determine if they display a similar organization. Also, it is crucial to be able to characterize quantitatively these features and to compare different cities with each other, eventually opening the possibility of a typology of cities. Another series of studies focused on obtaining a minimal set of measures by measuring the statistical significance of the correlations between various indicators \cite{Huang:2007, Schwarz:2010, Arribas:2011, Salvati:2016}. These papers consider indicators that characterize the urban form or the urban morphology, either by looking at the local distribution of population density or land use. However, these studies typically consider a large number of such indicators together with socio-economic indicators and leading to a clustering that is difficult to interpret. In particular, in \cite{Schwarz:2010} starting from a large number of indicators mixing urban form and socio-economical ingredients, six main variables are extracted. Using these variables, a classification of cities is proposed and leads in eight groups of cities that is however not based on the urban form only. 

In this paper, we will proceed in another way and we will specialize to the case of population density. Although our arguments could be extended to other measures, we concentrate on the problem of characterizing the spatial distribution of a single density, without aiming at understanding the interplay of population density with other socio-economic indicators. We do so by focusing on a dataset of a large number of french cities (about $4,500$ cities, of which $465$ cities have more than $10,000$ inhabitants) for which we have high-resolution data on the local population density. The density is obtained by covering the surface of the municipalities with a square grid of cells of size $200\times 200m^2$. The same grid is covering all the cities which enable us to measure various quantities without having to rely on artificial divisions in districts or patches that could introduce biases in the results.

We focus on the local population density defined as the density computed for a grid element, and we argue that the simplest characterization of this quantity has two main dimensions. The first one that was considered in many studies \cite{Tsai:2005,Schwarz:2010}, is the heterogeneity of the density distibution in a city and can be simply characterized by the corresponding Gini coefficient $G$. However this index is not able to characterize the other dimension that is the spatial organization of the city. We will focus on the spatial distribution of high density areas. Using a non-parametric method \cite{Louail:2014} we identify the large density regions, the population `hotspots', and characterize their spatial distribution with the use of a sprawling index $\eta$ computed as the normalized average distance between hotspots, and which tells us how dispersed in the city these hotspots are. Using these two dimensions, we are then able to cluster together similar cities and at an intermediate level, we construct four different types of cities.

\section{The Dataset}

In order to illustrate our theoretical discussion on a practical example we consider the local population density in cities in France. More precisely, the dataset considered is the population density covering a surface of about $58,000 \, km^2$, corresponding to the largest $10$ urban areas in France. The surface is covered with a square grid of $200 \times 200m^2$, and the population density is estimated for each grid cell. This dataset has been obtained by combining two different publicly available datasets: the INSEE dataset provides population for all French municipalities \cite{Insee:2015}, and living space data. We have combined these two datasets, by dividing the total population of each municipality on all the grid cells covering the surface of the municipality, proportionally to the living space area inside each cell. In this way we obtain an estimate of the number of inhabitants $P_i$ who reside in each grid cell $i$. Each grid cell $i$ belongs to one and only one municipality, and when a grid cell is on the border of two or more municipalities, it is assigned to the municipality with the largest surface share. We refer to the set of values $\{ i \in \alpha \}$ as the set of grid cells covering the municipality $\alpha$. We use latin indices such as $i, j$ for grid cells and greek indices such as $\alpha, \beta$ for municipalities (for example we will denote by $P_\alpha$ the population of city $\alpha$). Here we will use equivalently the population or the population density in cells, since all grids cells have the same area equal to $200\times 200=40,000 \, m^2$. We denote the population density in a grid cell by $\rho_i = P_i / A_i$, where $A_i$ is the area of grid cell.

\begin{figure}[ht!]
\centering
\includegraphics[scale=0.5]{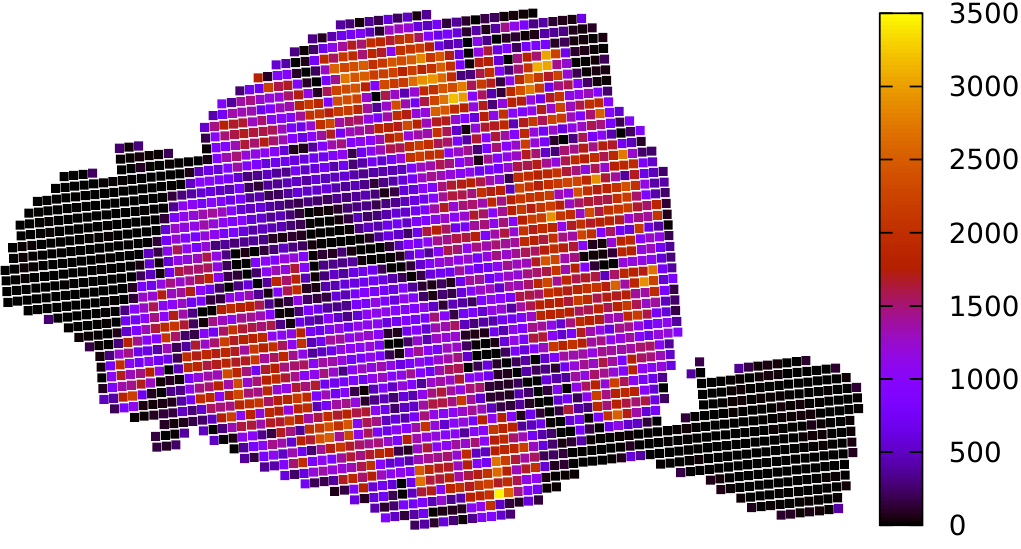}
\includegraphics[scale=0.5]{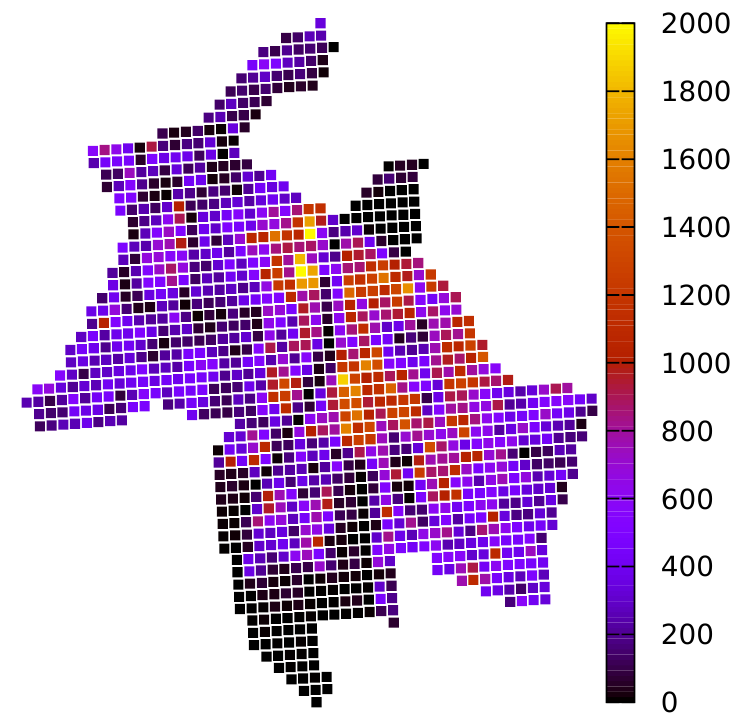}
\includegraphics[scale=0.5]{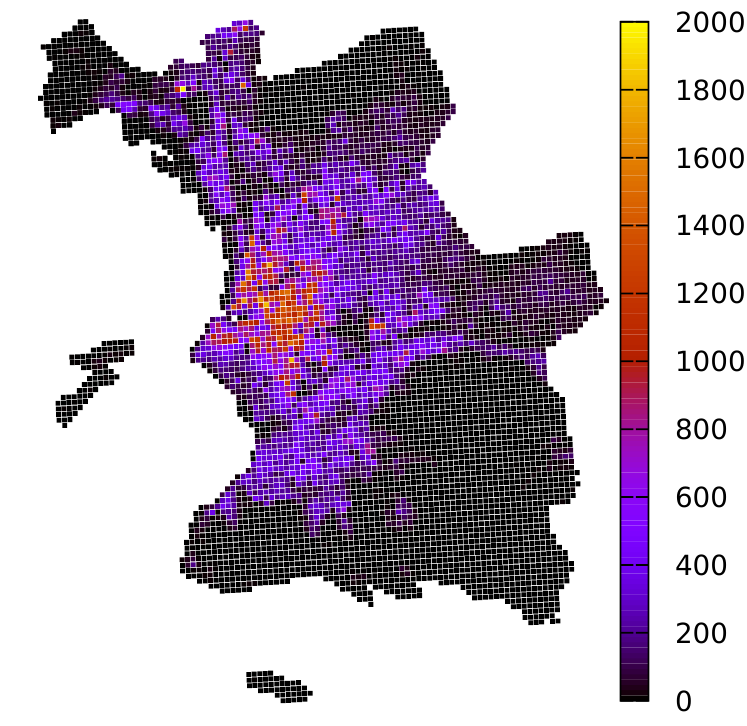}
\caption{\textbf{Population distribution the three largest french cities.} We represent here the local population density, obtained at the level of a square grid made of cells of size $200 \times 200$ meters. In the case of Paris we have $2,651$ grid cells, for Lyon $1,199$ cells, and for Marseille $6,533$ cells (all the numbers here are for 2013). The color of each grid cell refers to the estimated number of inhabitants per cell.}
\label{fig:fig1}
\end{figure}

\begin{figure}[ht!]
\centering
\includegraphics[scale=0.8]{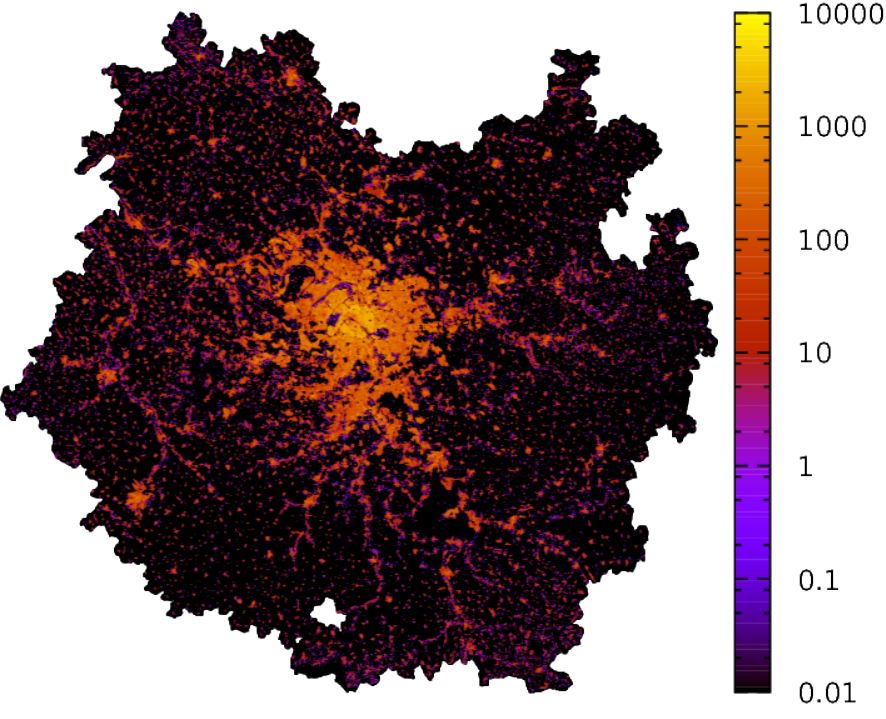}
\caption{\textbf{Population density in the whole Paris urban area.} The population density is obtained by covering cities with a square grid of $200 \times 200m^2$. Here, we display the example of the Paris urban area in 2013, covered by $592,714$ grid cells. The color of each grid cell refers to the estimated number of inhabitants per cell.}
\label{fig:fig2}
\end{figure}

The resolution in this dataset is enough to have a sufficiently clear picture of the population distribution for both the largest and the smallest municipalities. In Fig.~\ref{fig:fig1} we see some examples, by showing the population density spatial distribution at the scale of the individual municipality, for the three largest french cities: Paris, Lyon and Marseille, and in Fig.~\ref{fig:fig2} we display the distribution for the whole Paris urban area. 
We observe in general a monocentric pattern at the urban area scale, and a large variety of population distributions inside city boundaries. For example, in the case of Paris (France), the very center is mostly composed of offices and touristic places, leading to a low population density. In contrast, high density areas in Paris compose a ring-type area far from the geometrical center of the city. This is different from other cities such as Lyon or Marseille, where we observe a decreasing density with the distance to the barycenter. 

While the INSEE dataset contains the population of all french municipalities for the years 1968, 1975, 1982, 1990, 1999 and for all the years from 2006 to 2013, the living space data is available for 2008 and 2016 only. In this paper we then consider the population density at the grid cell level in 2008, and for 2013 we combined the INSEE population data for 2013 and the living space data for 2016. In the Table~\ref{tab:table1}, we present a summary of the dataset and in Figs.~\ref{fig:fig1}, \ref{fig:fig2} some example of maps obtained from it. 
\begin{table*}[ht!]
\small\sf\centering
\caption{\textbf{Summary of the dataset}. The largest 10 urban areas in France are covered by a square grid of cells of size $200 \times 200m^2$. For each urban area we list the total population for the years 2008 and 2013, the total number of grid cells covering the surface, and the number of municipalities existing in the urban area.}
\label{tab:table1}
\begin{tabular}{lllll}
\hline
Urban area & Total Pop. (2008) & Total Pop. (2013) & \# of cells & \# of municipalities \\
\hline
Paris & 12.6 M & 13.0 M & 592,714 & 2,320 \\
Lyon & 2.12 M & 2.24 M & 152,658 & 519 \\
Marseille & 1.75 M  & 1.77 M & 82,338 & 107 \\
Toulouse & 1.20 M  & 1.29 M & 137,131 & 453 \\
Lille & 1.21 M & 1.23 M & 25,210 & 132 \\
Bordeaux & 1.10 M & 1.17 M & 143,719 & 255 \\
Nice & 1.02 M & 1.02 M & 85,385 & 142 \\
Nantes & 0.85 M & 0.91 M & 85,582 & 114 \\
Strasbourg & 0.76 M  & 0.77 M & 56,414 & 267 \\
Rennes & 0.65 M & 0.70 M & 95,797 & 190 \\
\hline
\end{tabular}
\end{table*}

\section{Fluctuations of the local density}

\subsection{The Gini coefficient}

The first group of measures that we consider is designed to quantify how heterogenous the population density is inside the municipality boundaries, without taking into account the spatial nature of the data. These measures will depend on the set of population values of the grid cells only, and not on the spatial position of each cell. Amongst these measure, we consider the Gini index (or Gini coefficient) $G_\alpha$ for the municipality $\alpha$, defined as \cite{Dixon:1987}
\begin{equation}
G_\alpha = \frac{\sum_{i,j \in \alpha} | P_i - P_j |}{2 n_{\alpha} \sum_{i \in \alpha} P_i},
\end{equation}
where the sums run on all the $n_{\alpha}$ grid cells covering the surface of the municipality $\alpha$. The Gini coefficient is extensively used in Economics, and was originally introduced to quantify the levels of inequality in income and wealth distributions, but can be used for characterizing the level of heterogeneity of any quantity. In the city case, the Gini coefficient is designed to be zero for a uniform city in which the population is the same in all grid cells. On the contrary, the Gini is maximum for an extremely concentrated city, in which the total population resides in a single grid cell. In this case, the Gini, takes the value 
\begin{equation}
G_{\alpha}^{\ \textit{max}} = \frac{n_{\alpha} - 1}{n_{\alpha}},
\end{equation}
which is very close to $1$ for large $n_\alpha$. However, the Gini of municipalities covered by a very small number can be artificially small, even if the population is evenly distributed among the grid cells.
For this reason, we will use the normalized $G_{\alpha}^{\ \textit{norm}} $ defined as
\begin{equation}
G_{\alpha}^{\ \textit{norm}} = \frac{G_{\alpha}}{G_{\alpha}^{\ \textit{max}}} .
\end{equation}
Such normalized Gini is always defined between $0$ and $1$, and can be used to compare the Gini for different municipalities with very diverse numbers of cells. 

\begin{figure*}[ht!]
\centering
\includegraphics[scale=0.60]{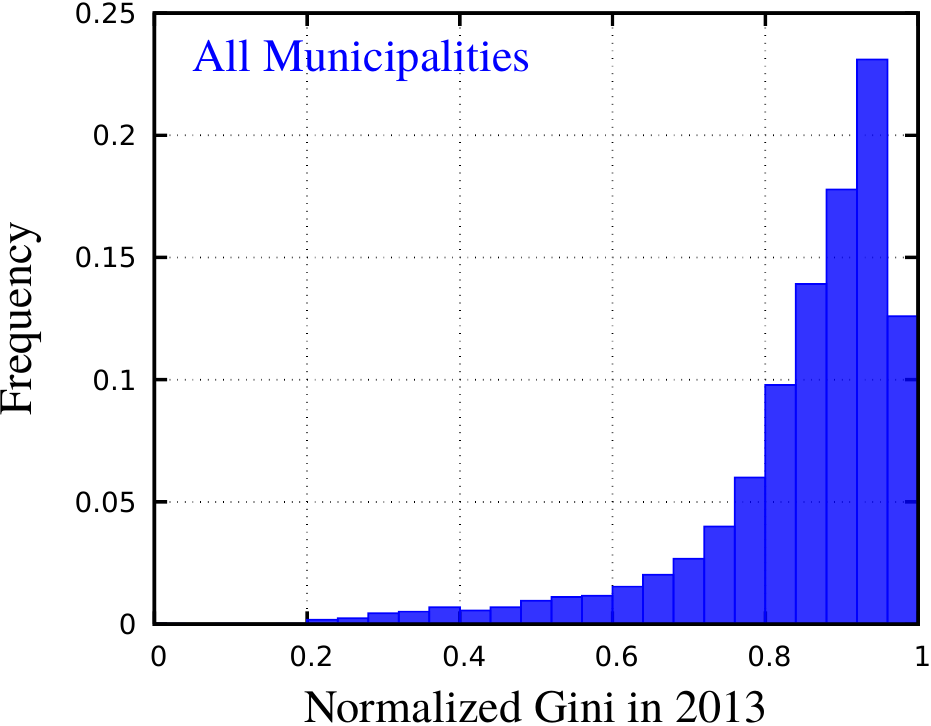}
\includegraphics[scale=0.60]{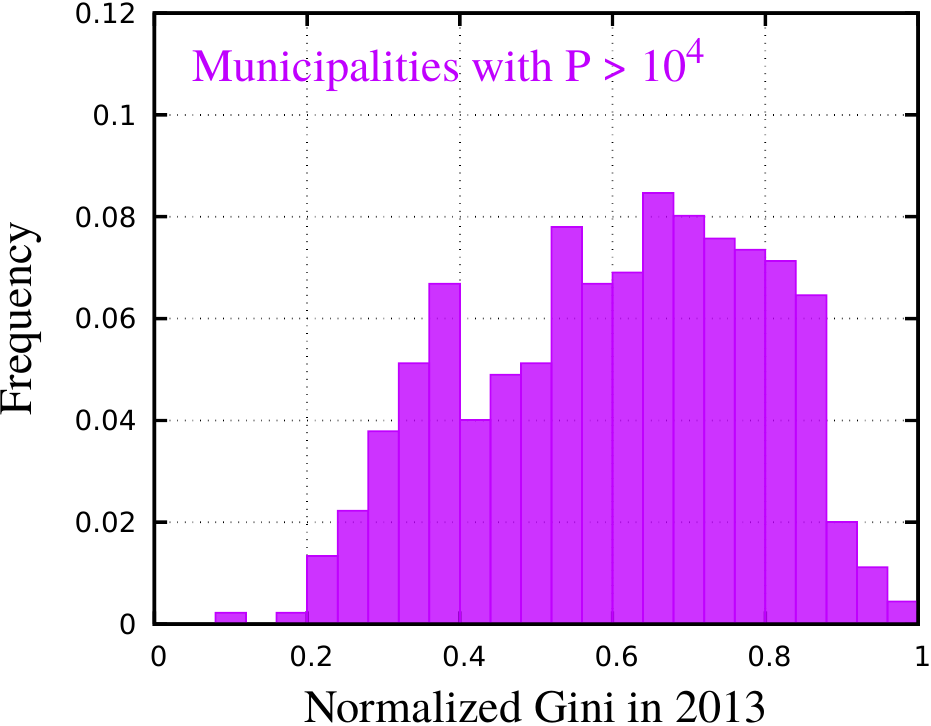}\\
\includegraphics[scale=0.60]{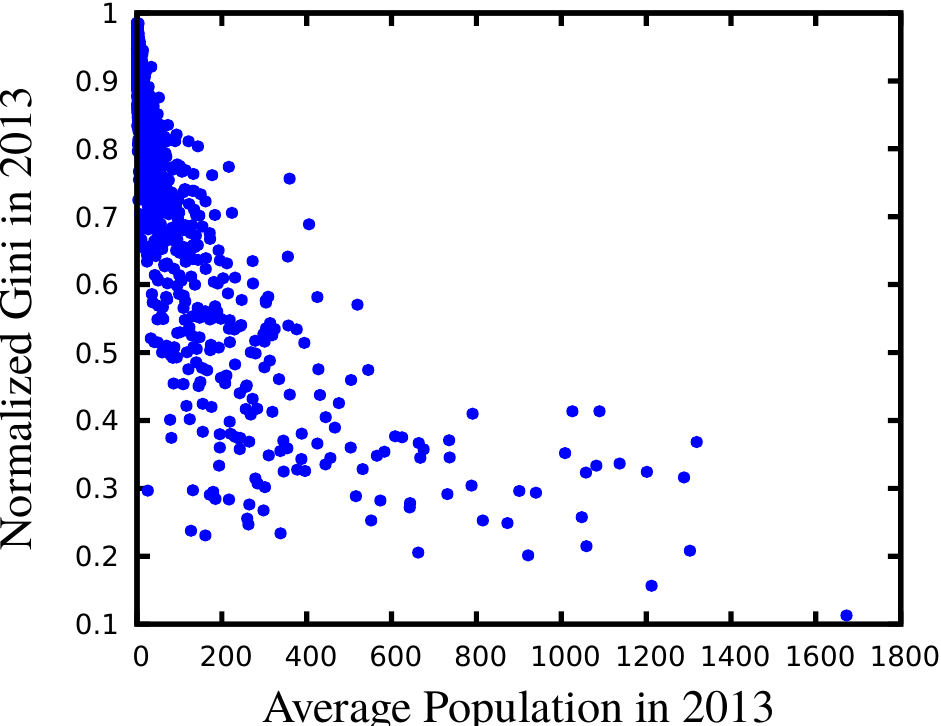}
\includegraphics[scale=0.60]{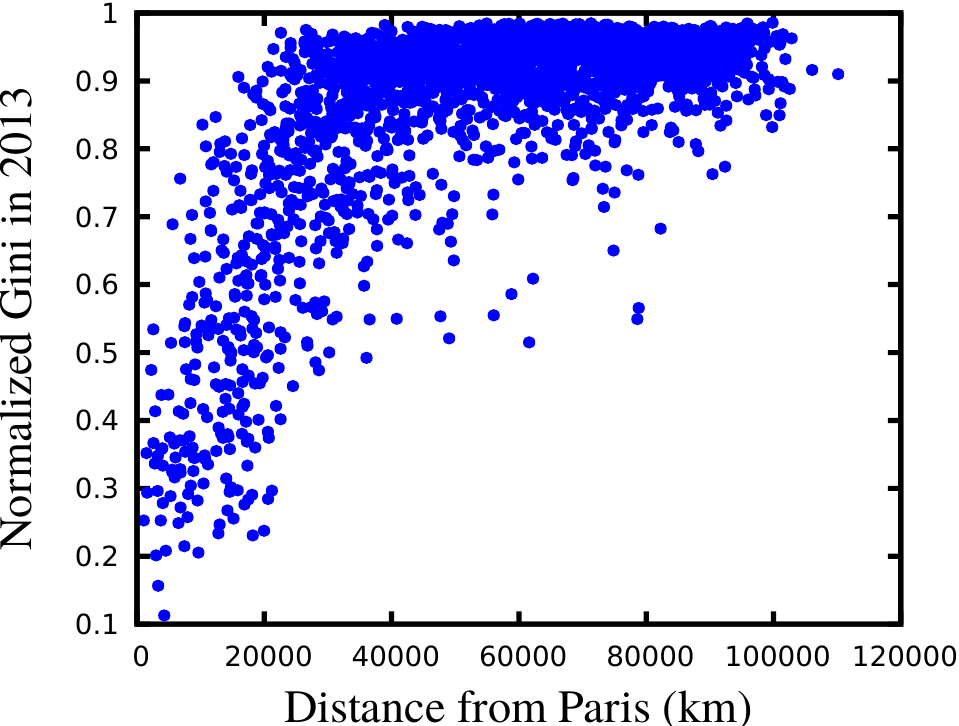}
\caption{\textbf{The normalized Gini coefficient in 2013.} The histogram shows the values of the normalized Gini coefficient for all municipalities considered in this study ($4,499$ municipalites). We notice that there is an abundance of municipalities with large values of Gini. However, if we focus on the $465$ municipalities with population larger then $10,000$ inhabitants, the distribution of the Gini is more uniform. In the bottom figures, we plot the normalized Gini coefficient for the municipalities in the Paris urban area versus their average population $\overline{P}$ in 2013 (left) and their distance from Paris (right). We see that the majority of municipalities with a large Gini have both low average population and are far from Paris.}
\label{fig:fig3}
\end{figure*}

In Fig. \ref{fig:fig3} we compute the Gini coefficient for all municipalities in our dataset in 2013. In the top left figure we notice that there is an abundance of municipalities with large values of Gini. Such an abundance can be explained by the heavy presence in the dataset of remote municipalities, that typically have a low average population and are far from the center of the urban area (see the bottom panels of Fig. \ref{fig:fig3}). If we focus on the municipalities with population larger than $10,000$ inhabitants, we observe a more uniform distribution of the Gini.

\begin{figure}[ht!]
\includegraphics[scale=0.7]{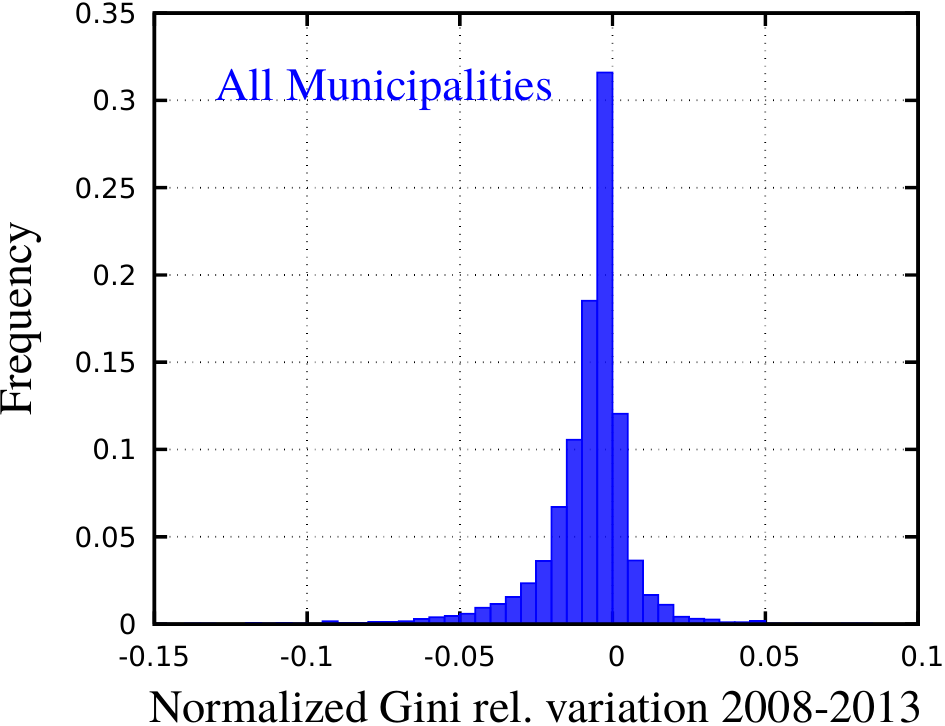}
\caption{\textbf{The relative difference between the normalized Gini in 2013 and in 2008}. The histogram represent the values of the normalized Gini relative variation ($G_\alpha(2013)-G_\alpha(2008)) / G_\alpha(2008)$ for all municipalities in the Paris urban area. We notice that for most of the municipalities ($3,592$ out of $4,499$, corresponding to $80\%$ of the municipalities) the Gini variation has been negative. The average value of this decrease is about $-0.77$ $\%$}
\label{fig:fig4}
\end{figure}

We also have considered the time variation of the normalized Gini coefficient (see Fig. \ref{fig:fig4}). In the time period considered, between 2008 and 2013, the normalized Gini has been reducing for about 80$\%$ of the municipalities. This trend is in line with the relation between the Gini and the average population, showed in Fig.~\ref{fig:fig3}. As urbanization increases, the average population increases, which results in a reduction of the Gini coefficient for most of the municipalities. Such variations in the normalized Gini coefficient are however typically small, the average normalized Gini relative variation is about $-0.77\%$, and they are of the order of few percent for all municipalities.

\subsection{Alternative measures: Standard deviation and entropy}

Alternative measures of statistical dispersion can be found in the literature and the most common ones are the standard deviation and the entropy (in a urban context, see for example \cite{Tsai:2005} and references therein). These measures can be defined for every municipality in the dataset, and similarly to the Gini coefficient, they depend on the population distribution $\{ P_i \}$, but not on their spatial distribution. In particular, the relative standard deviation is defined as
\begin{equation}
\sigma^{\ \textit{rel}}_{\alpha} = \left( \overline{P^2}_{\alpha} - \overline{P}^2_{\alpha} \right)^{1/2}/\overline{P}_{\alpha},
\end{equation}
where $\left( \overline{P^2}_{\alpha} - \overline{P}^2_{\alpha} \right)^{1/2}$ and $\overline{P}_{\alpha}$ are respectively the standard deviation and the average of the distribution of $P_{\alpha} = \{P_i, \ \forall i \in \alpha\}$. The entropy of the distribution is defined as \cite{Cover:2012}
\begin{equation}
S_{\alpha} = - \frac{1}{\log n_{\alpha}} \sum_{i \in \alpha} p_i \log(p_i)
\end{equation}
where $p_i$ is the normalized population distribution $p_i = P_i / (n_\alpha \overline{P}_{\alpha})$.

\begin{figure}[ht!]
\includegraphics[scale=0.60]{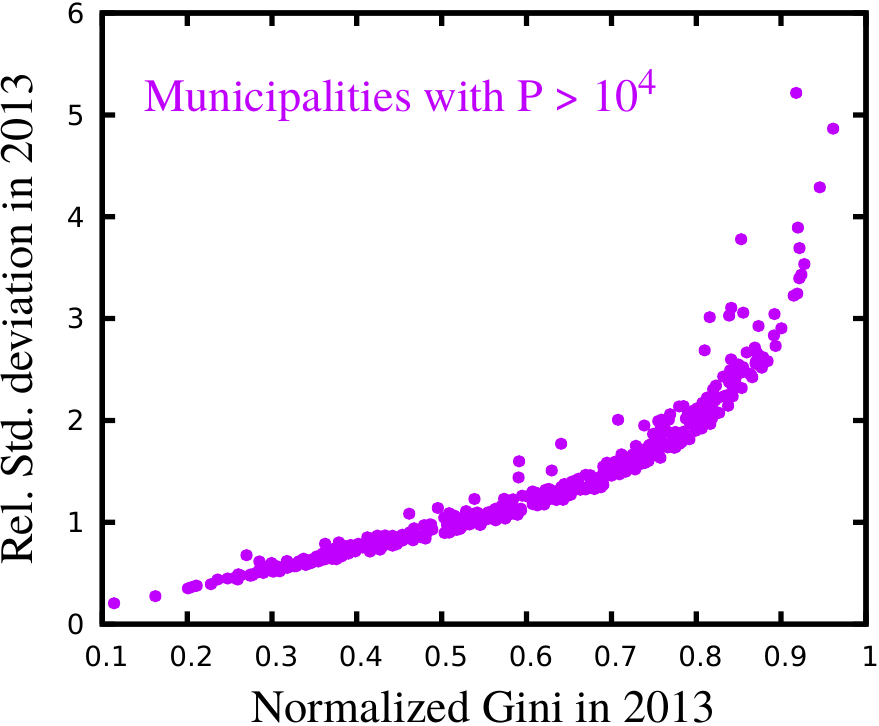}
\includegraphics[scale=0.60]{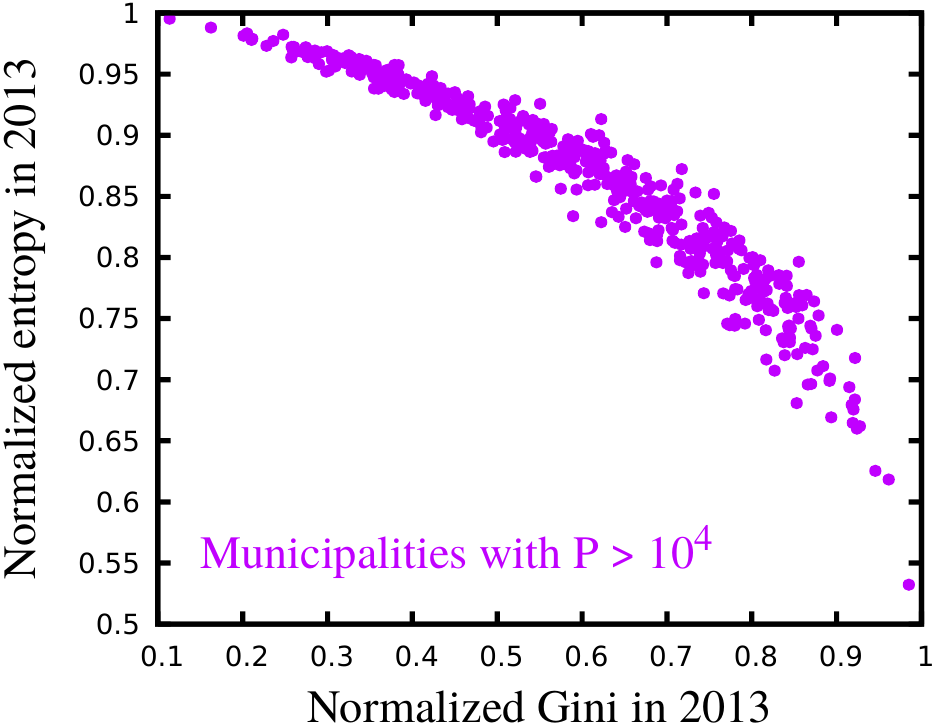}
\caption{\textbf{The Gini coefficient and other measures of statistical dispersion}. We plot here the normalized Gini coefficient versus the relative standard deviation (left) and the entropy (right), for the $465$ municipalities in our dataset with population larger then $10,000$ inhabitants (in 2013). We excluded small cities, that are typically covered by a smaller number of grids in order to reduce the noise in the plots. The monotonic behaviour shows that cities with large Gini have also large relative standard deviation and low entropy, meaning that different measures of dispersions contain essentially the same information.}
\label{fig:fig5}
\end{figure}

However, all these measures are equivalent to the Gini coefficient. In Fig.~\ref{fig:fig5} we plot for all the municipalities in our dataset in 2013, the Gini and different measures of statistical dispersions. The monotonic behaviour shows that there is a bijection between the Gini and these measures, demonstrating that they convey the same information about the population distribution in the city. Using the Gini or these measures for characterizing the heterogeneity of the population would then lead to the same results. 

\section{The spatial organization of large density locations}

\subsection{Gini is not enough}

\begin{figure}[ht!]
\centering
\includegraphics[scale=0.4]{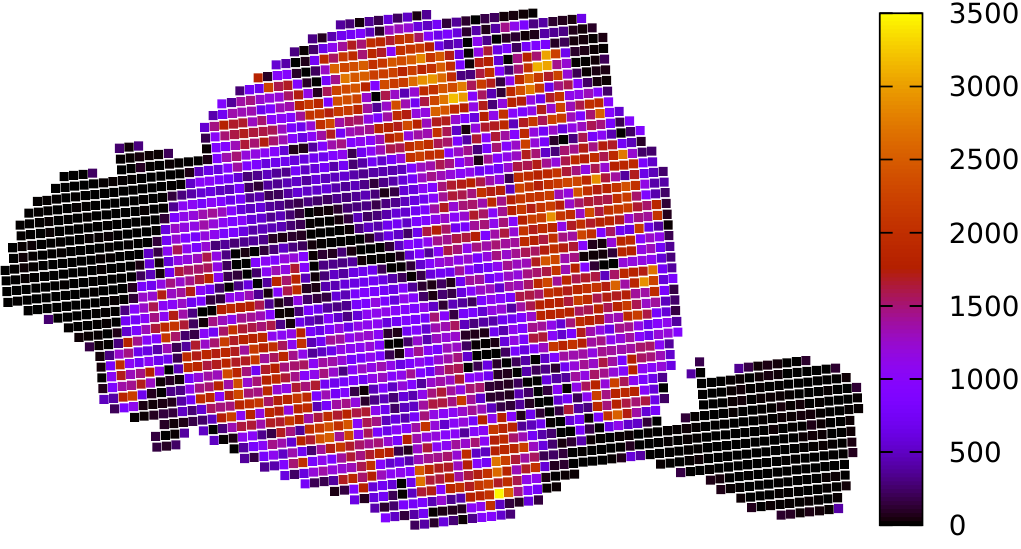}
\includegraphics[scale=0.4]{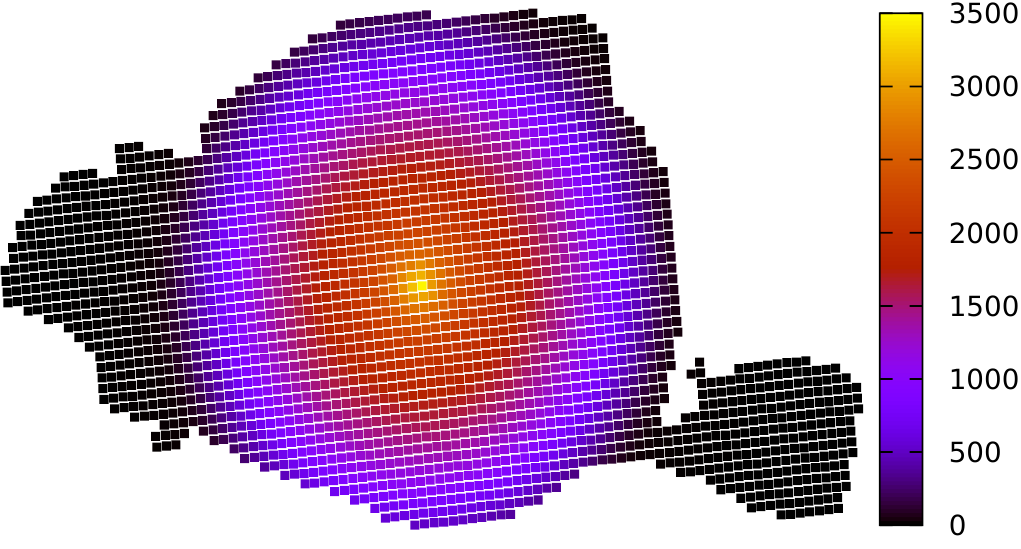}\\
\includegraphics[scale=0.4]{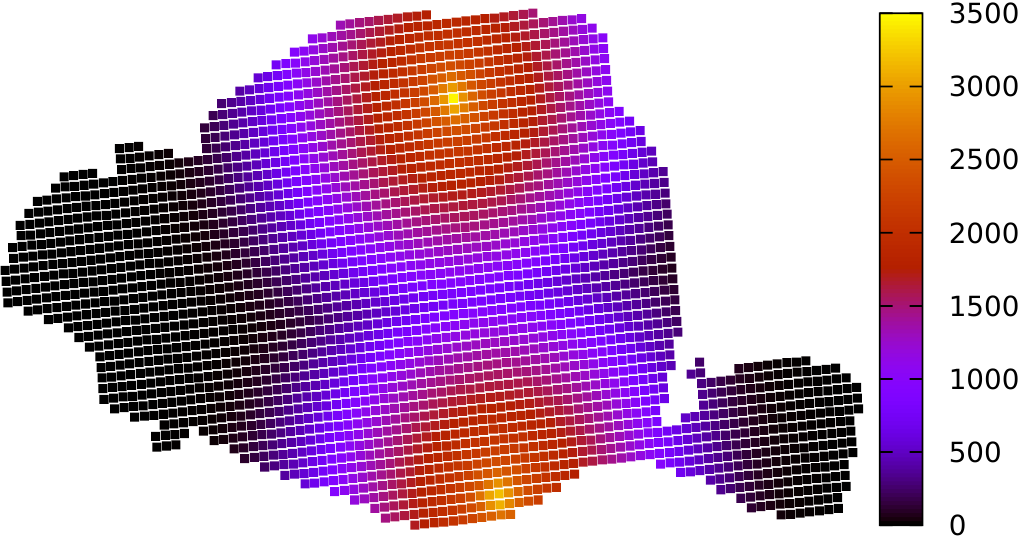}
\includegraphics[scale=0.4]{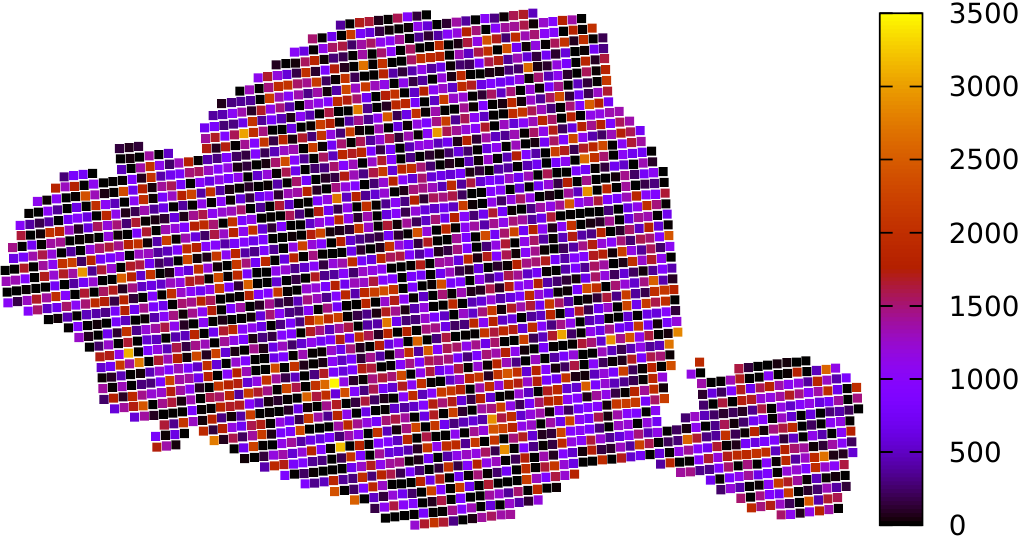}
\caption{\textbf{The population distribution in Paris in 2013 and different distributions with exactly the same Gini}. The different population distributions are constructed by just reshuffling the values of the population, obtained from the real population density distribution (top left). We can then construct an ideal monocentric city, an ideal polycentric city (with two centers) or a completely random city. Even though the spatial distribution in the four maps are very different, the Gini coefficient (and other measure of statistical dispersion), is the same for all maps $G \simeq 0.49$, since it does not depend on the spatial distribution of population.}
\label{fig:fig6}
\end{figure}

As mentioned in the previous section, quantifying heterogeneities through measures of statistical dispersions such as the Gini coefficient, provides interesting measures that however are not affected by how the grid cells are distributed in space. In Fig.~\ref{fig:fig6} we show this explicitly in the case of the city of Paris: Spatial reshuffling of the population values among the different grid cells covering the same surface, do not affect the Gini.

At this point, once we have quantified the population heterogeneity, we have to characterize how it is organized in space. In the case  of low heterogeneity, most grid cells have roughly the same population and the spatial organization is largely irrelevant. Space becomes relevant when we have large population fluctuations, with the appearance of empty cells and very densely populated areas, that we will also denote in the following `hotspots'. How population is spatially arranged in the city is then largely determined by the locations of these hotspots, hence in the following we will focus on these objects. 

\subsection{The hotspots}

\begin{figure}[ht!]
\includegraphics[scale=0.4]{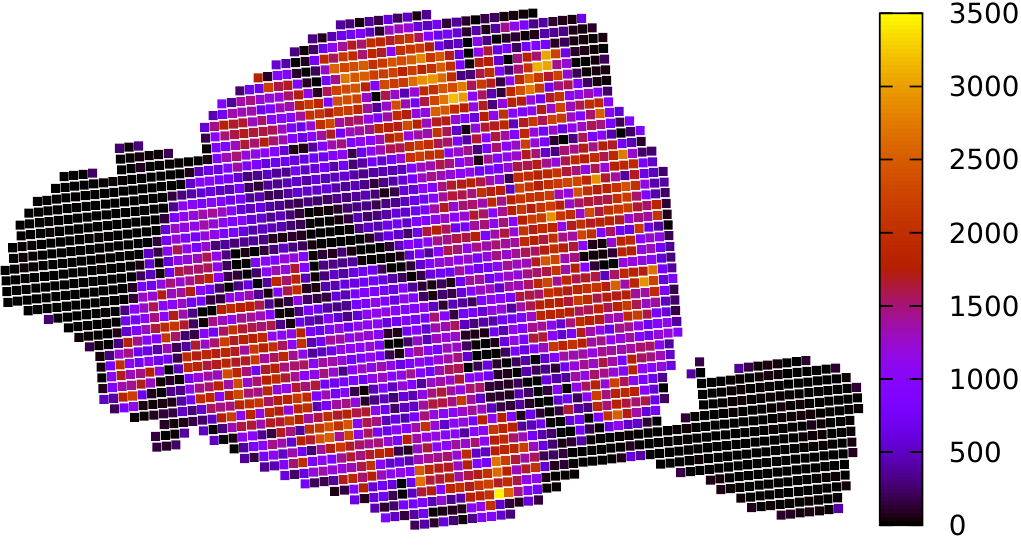}
\includegraphics[scale=0.4]{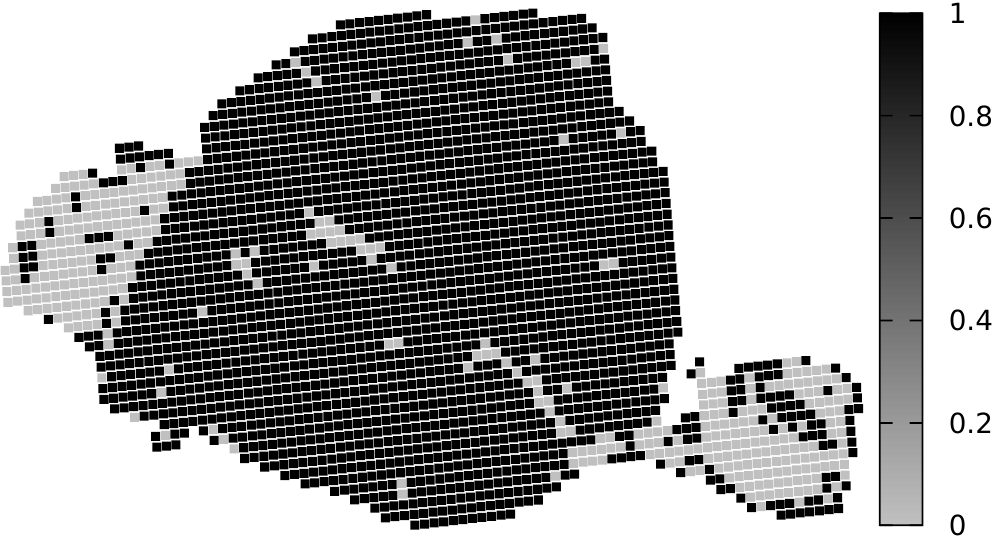}\\
\includegraphics[scale=0.4]{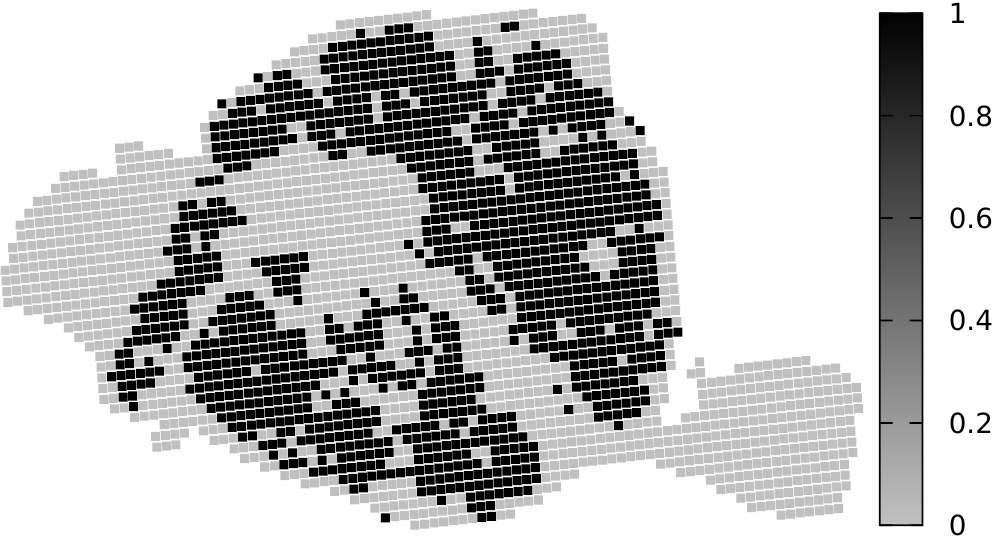}
\includegraphics[scale=0.4]{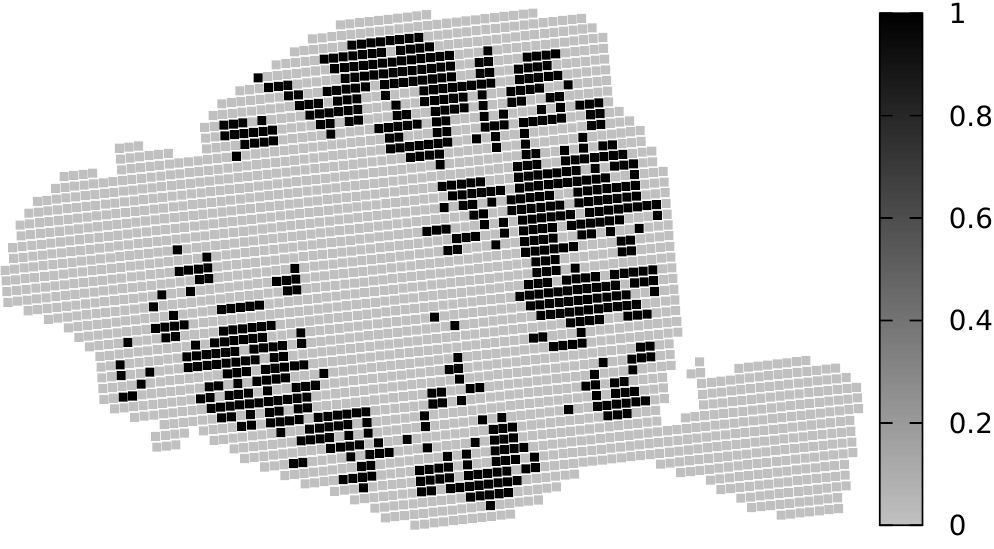}
\caption{\textbf{The population distribution in Paris in 2013 (top left) and the maps obtained considering different population thresholds}. (top right) The grid cells with more than $1$ inhabitant per cell are highlighted in black, (bottom left) the grid cells with population larger than the average population (of Paris) per cell are selected, (bottom right) the grid cells with population larger then the LouBar population value (see the main text and \cite{Louail:2014}).}
\label{fig:fig7}
\end{figure}

As discussed above, we argue that the spatial arrangement is almost entirely characterized by the locations of high density areas, the hotspots. These hotspots are defined as the grid cells with a population above a certain threshold $P^*$ and we illustrate this idea in Fig.~\ref{fig:fig7}. We vary the value of the threshold in the case of Paris, and display only the grid cells with population above it. This sequence allows us to locate the most important population hotspots in the french capital.

\begin{figure}[htbp!]
\centering
\includegraphics[scale=0.8]{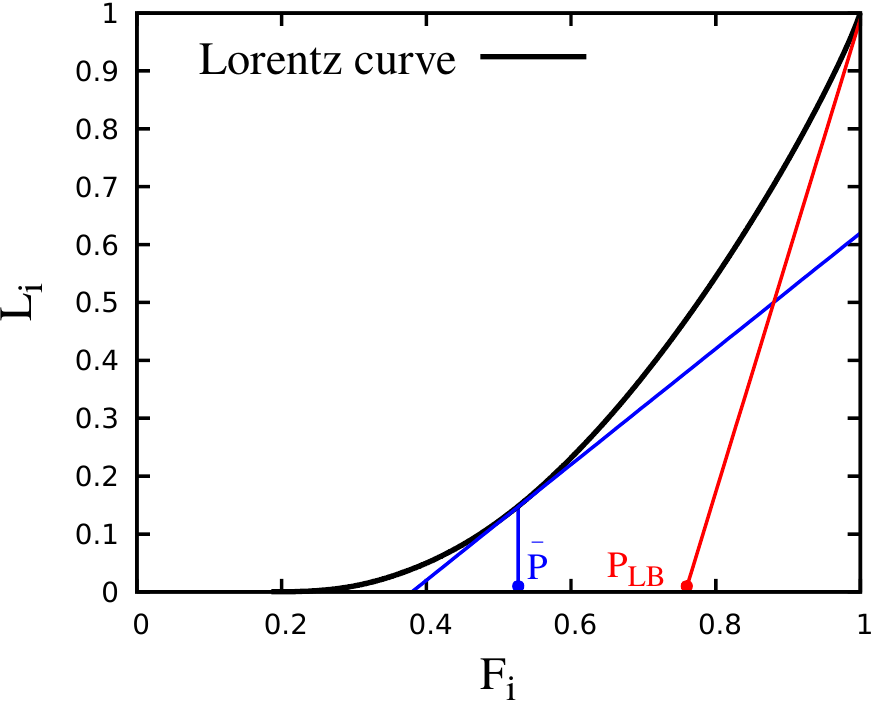}
\caption{\textbf{The Lorenz curve for the population in Paris in 2013}. For any distribution, the Lorenz curve can be constructed in the following way. For a set of values of size $n$, we order the sequence of values $P_{i}$ with $i = 1 \cdots n$ in non-decreasing order. We plot the value of the incomplete sums $L_i \equiv \left( \sum_{j=1}^{i} P_i \right) / \left( \sum_{j=1}^{n} P_i \right) $ versus $F_i \equiv i / n$. Interestingly, the average of the $P_i$ can be obtained by the projection $F_P$ on the $x$-axis of the tangent of slope $1$ and by inverting $F(\overline{P})=F_P$. The $P_{LB}$ value is found from the  intersection of the $x-$axis with the tangent of the Lorenz curve at $F_i = 1$ (red line). For Paris, the average population $\overline{P}$ is around 847 inhabitants while the value $P_{LB}$ is around $1,462$ inhabitants. The number of cells with population larger than the average are $1,254$ over $2,651$, while the number of cells with population larger then the Loubar value are $649$. }
\label{fig:fig8}
\end{figure}

The choice of the threshold defining the hotspots is therefore critical and we will use here a non-parametric method for choosing this threshold, introduced previously in \cite{Louail:2014}. In principle, different threshold are possible, such as for instance the average population $\overline{P}$, which is the most naive choice, and the LouBar population value $P_{\textit{LB}}$ that we are going to use here. These threshold can be found using the Lorenz curve, shown for the population distribution of Paris in Fig.~\ref{fig:fig8}. In the caption of Fig.~\ref{fig:fig8} we give a brief description of how to select the non-parametric value $P_{\textit{LB}}$.

Once we have determined the hotspots located where the grid cells have a population larger than a threshold $P^*$, we can compute the average distance between all of them and compare this distance to some measure of the size of the city. This leads us to define the \emph{spreading index at level} $P^*$ defined as
\begin{equation}
\eta_{\alpha} (P^*) = \frac{ \frac{1}{n_\alpha (P^*)} \sum_{i, j} d(i,j) \Theta(P_i - P^*) \Theta(P_j - P^*)}{\frac{1}{n_\alpha}\sum_{i, j} d(i,j)} ,
\end{equation}
where $n_\alpha (P^*)$ is the number of municipalities with population larger then $P^*$ and  $d(i,j)$ is the distance between the grid cell $i$ and the grid cell $j$. The sums both in the numerator and the denominator are performed over all the grid cells covering the surface of municipality $\alpha$. While the denominator is the average distance between all the grid cells, irrespective of their population and gives a measure of the city size, the numerator is the average distance between all the grid cells for which the population is above a certain threshold, $P_{i (j)}  > P^*$. If the large populated areas (the hotspots), are spread all around the surface of the municipality, this ratio will be large (close to $1$). On the contrary, if the hotspots are close to each other, as in the case of a monocentric city, this ratio will be small. 

\begin{figure}[ht!]
\includegraphics[scale=0.7]{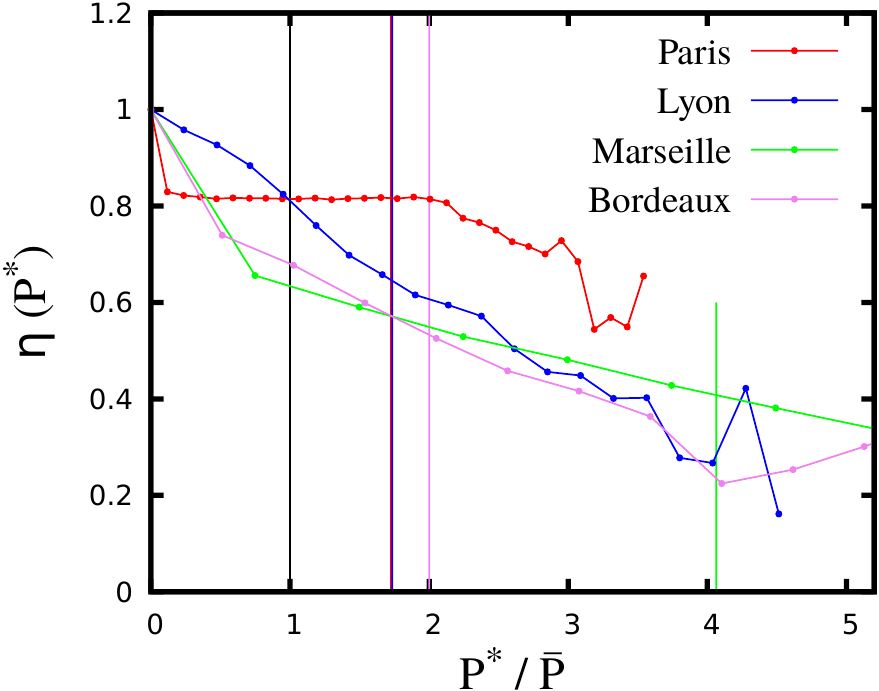}
\includegraphics[scale=0.7]{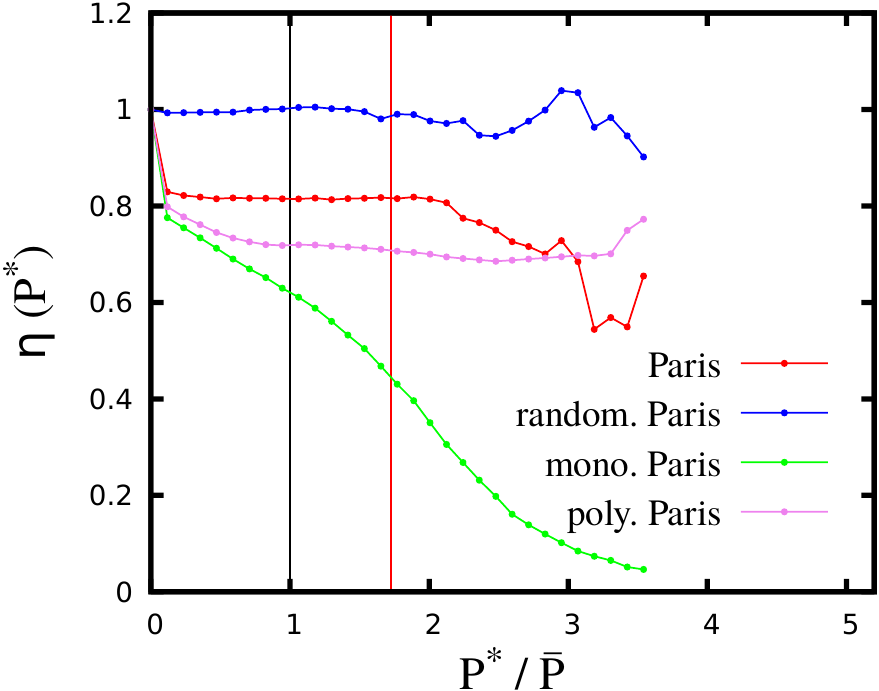}
\caption{\textbf{The spreading index for different values of thresholds}. We show the spreading index as a function of the threshold population $P^*$, rescaling the curves versus $P^*/\overline{P}$. The vertical lines above $1$ corresponds to the LouBar values. We plot these curves for (left) four french cities and (right) the reshuffling of Paris population density distributions showed in Fig.~\ref{fig:fig6}.  The decreasing behaviour of $\eta(P^*)$ as a function of the threshold $P^*$ is a signature of monocentricity. It is visible in most of large cities, except for Paris.}
\label{fig:fig9}
\end{figure}

In Fig. ~\ref{fig:fig9}, we show the variation of the spreading index when we vary the threshold $P^*$. 
We plot this curve for four french cities (left), and also for the reshuffled densities of Paris shown in Fig.~\ref{fig:fig6}. In particular, we observe that monocentricity is characterized by a monotonic decrease of $\eta(P^*)$ with $P^*$ (for example Lyon, Marseille, Bordeaux and the monocentric reshuffling of Paris ). In contrast for non-monocentric cities, the behavior is more complex with the appearance of plateaux.  We notice that for the value of the threshold $P^* = \overline{P}$ the spreading index for Paris is approximately equal to the spreading index for Lyon. This seems in contrast with the observed differences between the two cities. 

\begin{figure*}[ht!]
\centering
\includegraphics[scale=1.0]{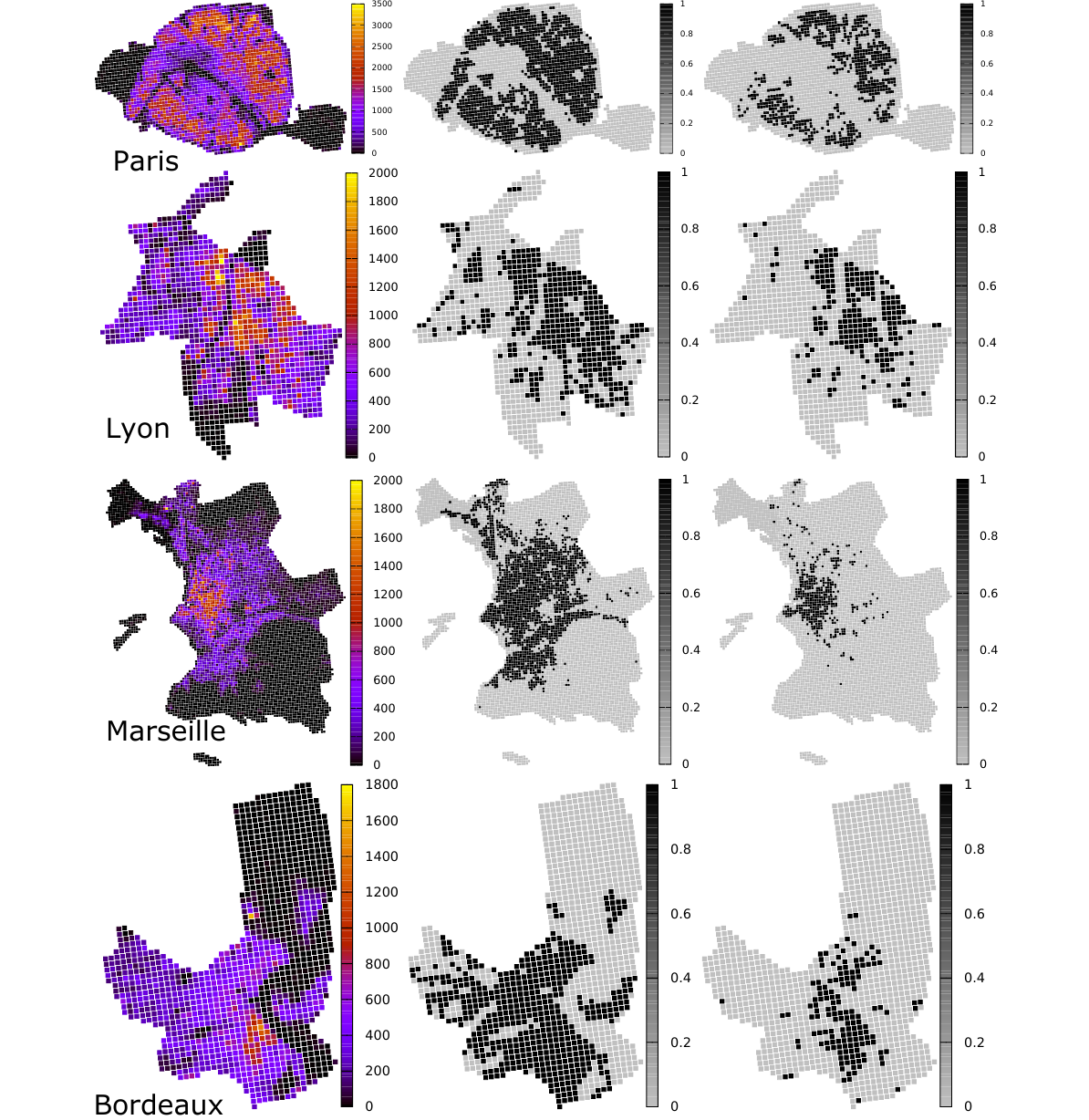}
\caption{\textbf{Maps and hotspots for different thresholds for four major cities in France.} We show here the maps of Paris, Lyon, Marseille and Bordeaux and the hotspots maps, using the average value (1st column) and the LouBar value (2nd column) as a population threshold for identifying hotspots. We observe that the  more restrictive value $P_{\textit{LB}}$  allows to identify more clearly differences between cities.}
\label{fig:fig10}
\end{figure*}

In the following we will essentially consider the threshold values $P_{\textit{LB}}$ and $\overline{P}$ for computing the spreading index. In Fig.~\ref{fig:fig10} we show the maps of the four french cities presented above, and their hotspots maps for these two values of the threshold $P^*$. In the case $P^*=\overline{P}$, approximately half of the grid cells are considered hotspots, and the differences between population spatial distributions are not visible. On the contrary, with the LouBar value the characteristics of the difference cities become visible, in particular we observe in the case of Paris that large populated areas display a ring-type pattern around the geometrical center of the city, while others cities show clearly a monocentric distribution. It seems therefore more sensible to use the more restrictive threshold $P_{\textit{LB}}$ and for example we see in Fig~\ref{fig:fig9} that if we use $\eta(P_{\textit{LB}})$ to measures compactness/spraedness of cities, Paris is more spread than Lyon, which is more spread than Bordeaux, and more spread than Marseille, confirming the visual impression given by Fig.~\ref{fig:fig10}.

\section{The two dimensions of the population density}

\subsection{The average behavior}

\begin{figure}[ht!]
\centering
\includegraphics[width=0.4\textwidth]{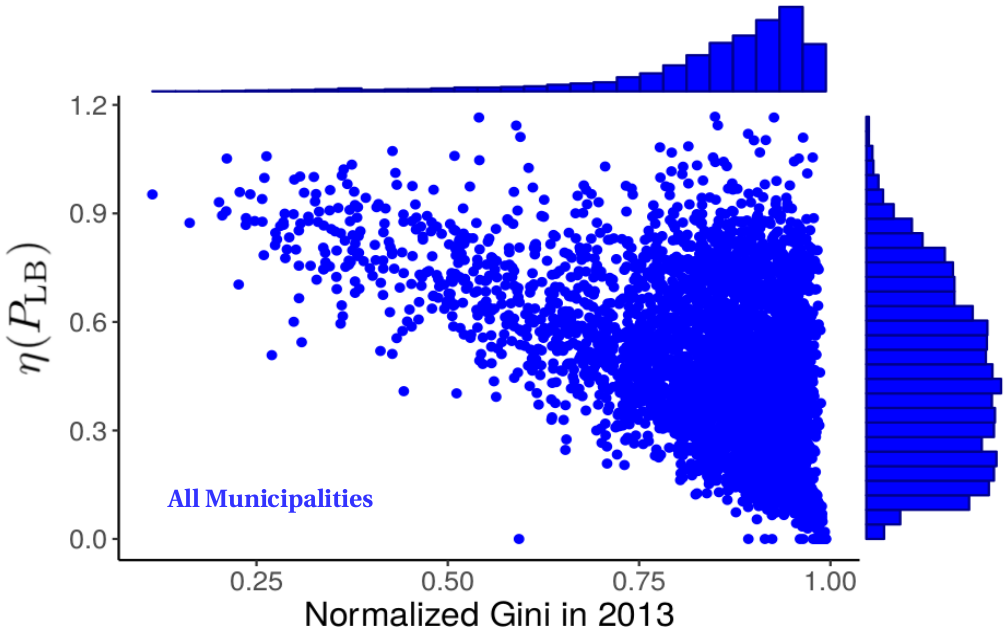}
\includegraphics[width=0.4\textwidth]{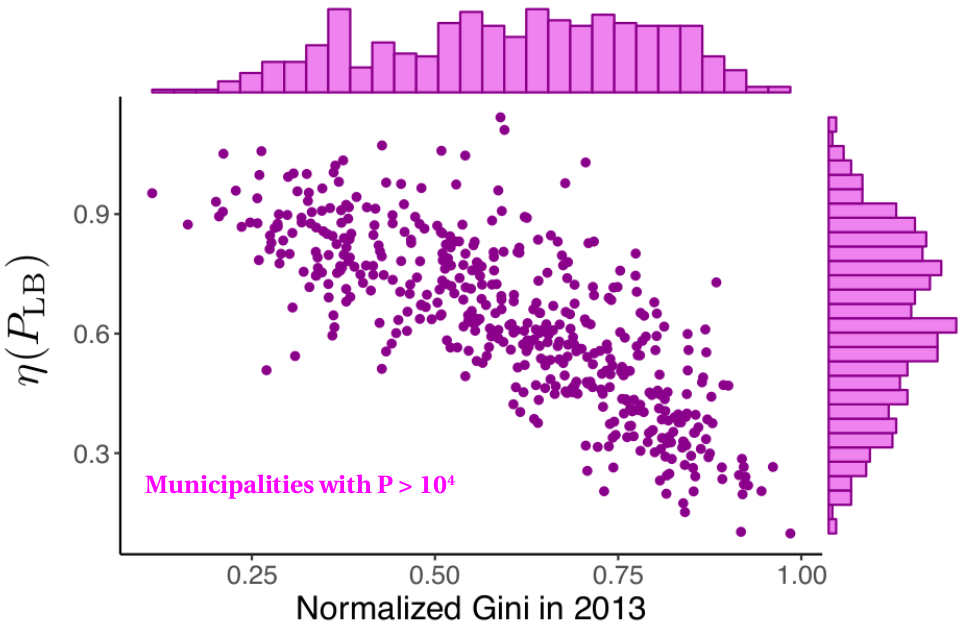}
\caption{\textbf{ The spreading index $\eta (P_{\text{LB}})$ versus the normalized Gini coefficient}. We show here all the municipalities of our dataset in 2013 on the plane $(G_\alpha, \eta_\alpha)$. In the left figure we show all the municipalities, while on the right we show only the municipalities with more than $10,000$ inhabitants. When small municipalities are excluded, a negative correlation between spreading index and Gini coefficient is revealed, with large fluctuations around this average trend.}
\label{fig:fig11}
\end{figure}

After having defined $\eta(P_{\textit{LB}})$, an index that characterizes the degree of compactness of highly populated areas, we can combine its information with the one contained in the Gini coefficient. In Fig.~\ref{fig:fig11} (left) we plot the two quantities for the municipalities in our dataset. As discussed before, if we plot all cities we notice an abundance of municipalities with large value of Gini. These municipalities are typically small, with low average population density and located far from the center of the municipalities. A more interesting analysis can be conducted by focusing on cities that are large enough with population larger than $10,000$ inhabitants. In Fig.~\ref{fig:fig11}(right), we observe for these large cities that on average the spreading index decreases with the Gini coefficient: cities with large (small) values of the Gini are city where the spreading index is rather small (large). This implies that typically a city with large population heterogeneity will be more compact. The two quantities do however contain a different information, as we can see from the scatter plot:  we observe homogeneous and heterogeneous cities in which the population distribution is either localized or delocalized.

\subsection{Clustering: four types of cities}

We will use here a hierarchical clustering method \cite{Friedman:2001} applied to cities and using their location in the 
$(G,\eta)$ plane. Hierarchical clustering proceeds by aggregating a set of data, starting from an initial configuration, in which each point is considered as belonging to a different cluster, and the number of cluster is the same as the number of points. The algorithm builds the hierarchy by progressively merging clusters step by step, by putting together clusters which are the closest. In our case, the standard euclidean metric distance in the space $(G, \eta)$ is used to compute distances:
\begin{equation}
d(\alpha,\beta)=\left[(G_\alpha-G_\beta)^2+(\eta_\alpha-\eta_\beta)^2\right]^{1/2}
\end{equation}
When the distance between two clusters needs to be computed, different choices can be made. We use here a complete-linkage clustering \cite{Defays:1977}, which define the distance between two clusters as the maximum distance between all the possible distance that can be considered between all the points in the first cluster and all the points in the second cluster. Complete-linkage clustering is preferrable because penalizes the merging of large clusters, and it tends to obtain clusters of approximately equal sizes \cite{Defays:1977}. On the contrary, others methods, such as the single-linkage clustering or the average-linkage clustering, often ends up with a very unequal cluster size distribution, with most of the points of the dataset belonging to a single cluster.

\begin{figure}[htbp!]
\centering
\includegraphics[width=0.5\textwidth]{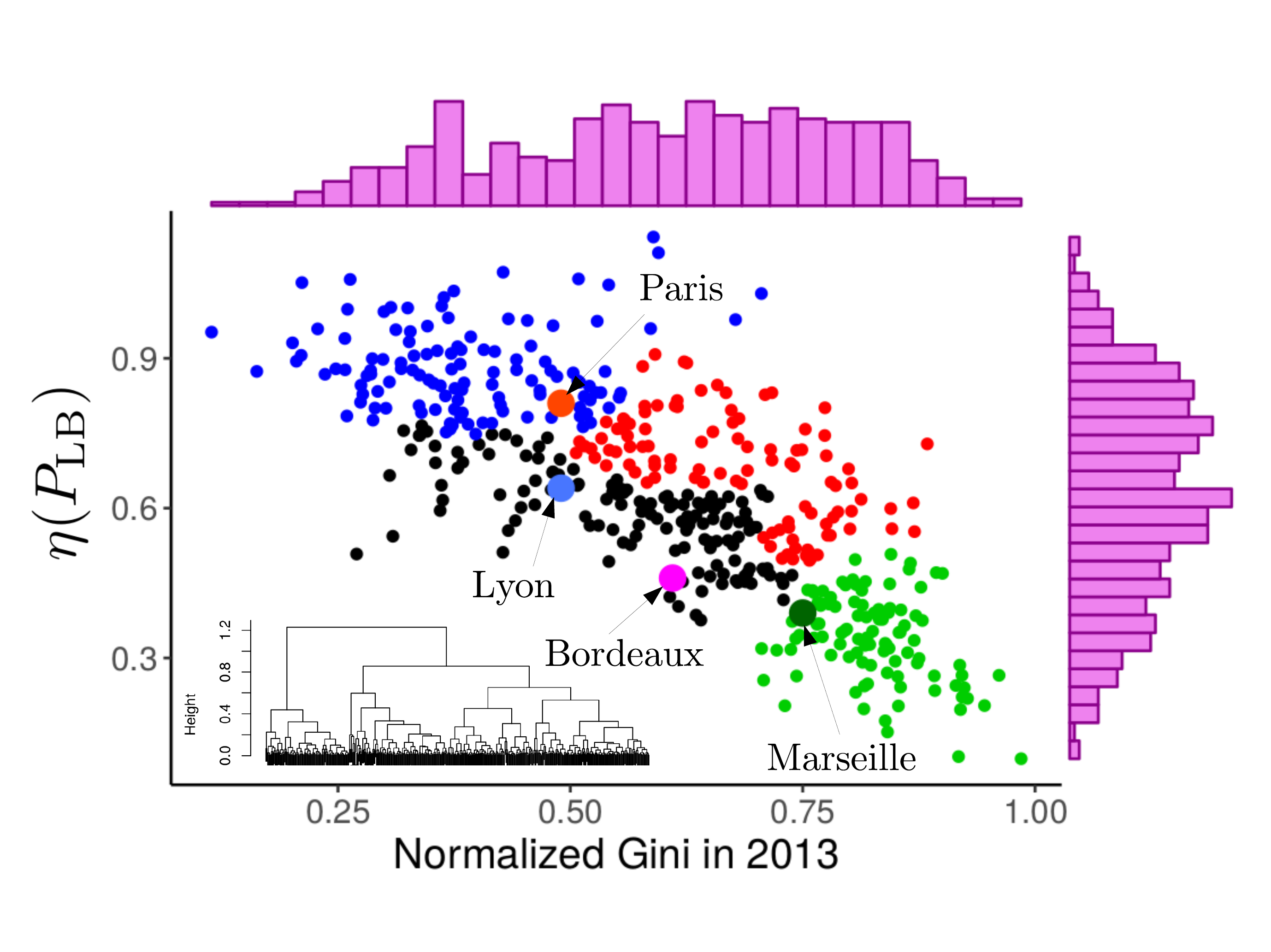}
\caption{\textbf{ Clustering cities using the spreading index $\eta (P_{\text{LB}})$ and the Gini index} We plot here all municipalities with more than $10,000$ inhabitants, and we apply a hierarchical clustering algorithm. At an intermediate level of the dendrogram shown here, we identify four clusters: (i) in blue, homogeneous and dispersed cities where the density fluctuations are small, (ii) in green, very heterogeneous cities with a compact organization of large densities areas. The last two groups comprise heterogeneous cities with (iii) a monocentric organization (in black) or (iv)  a more delocalized, polycentric structure (in red).}
\label{fig:fig12}
\end{figure}

In Fig.~\ref{fig:fig12} we see the result of the clustering algorithm, and we show here four clusters obtained at an intermediate level of the dendrogram. Due to the negative correlation between $G$ and $\eta$, the clustering divides cities along the direction in which lies the scatter cloud. We observe a first family of homogeneous cities (in blue), in which population density heterogeneities are negligible. We also observe (In green) cities in which the population distribution is very heterogeneous, and typically they are also very compact (small values of $\eta$). In the intermediate phase, we have the two clusters of cities (in black and red) which are neither too much homogeneous and spread, nor too heterogeneous and compact. We see that at this level the city of Paris belong to the first cluster (homogeneous and spread), the city of Marseille to the second cluster (very heterogeneous and compact) and the cities of Lyon and Bordeaux are in the intermediate clusters.

Furthermore, we see what the clustering algorithm divides the intermediate region in two clusters within the same range of $G$, nd approximately along the line describing the linear relation between $G$ and $\eta$. The cluster in red represents cities that are more spread than the average (for this level of heterogeneities) and the black cluster contains cities that are more compact than one would guess on average from their heterogeneity level. Lyon and Bordeaux both belong to the latter cluster, due to their monocentrical structure, see Fig.~\ref{fig:fig9}. 

\begin{figure}[htbp!]
\centering
\includegraphics[width=0.50\textwidth]{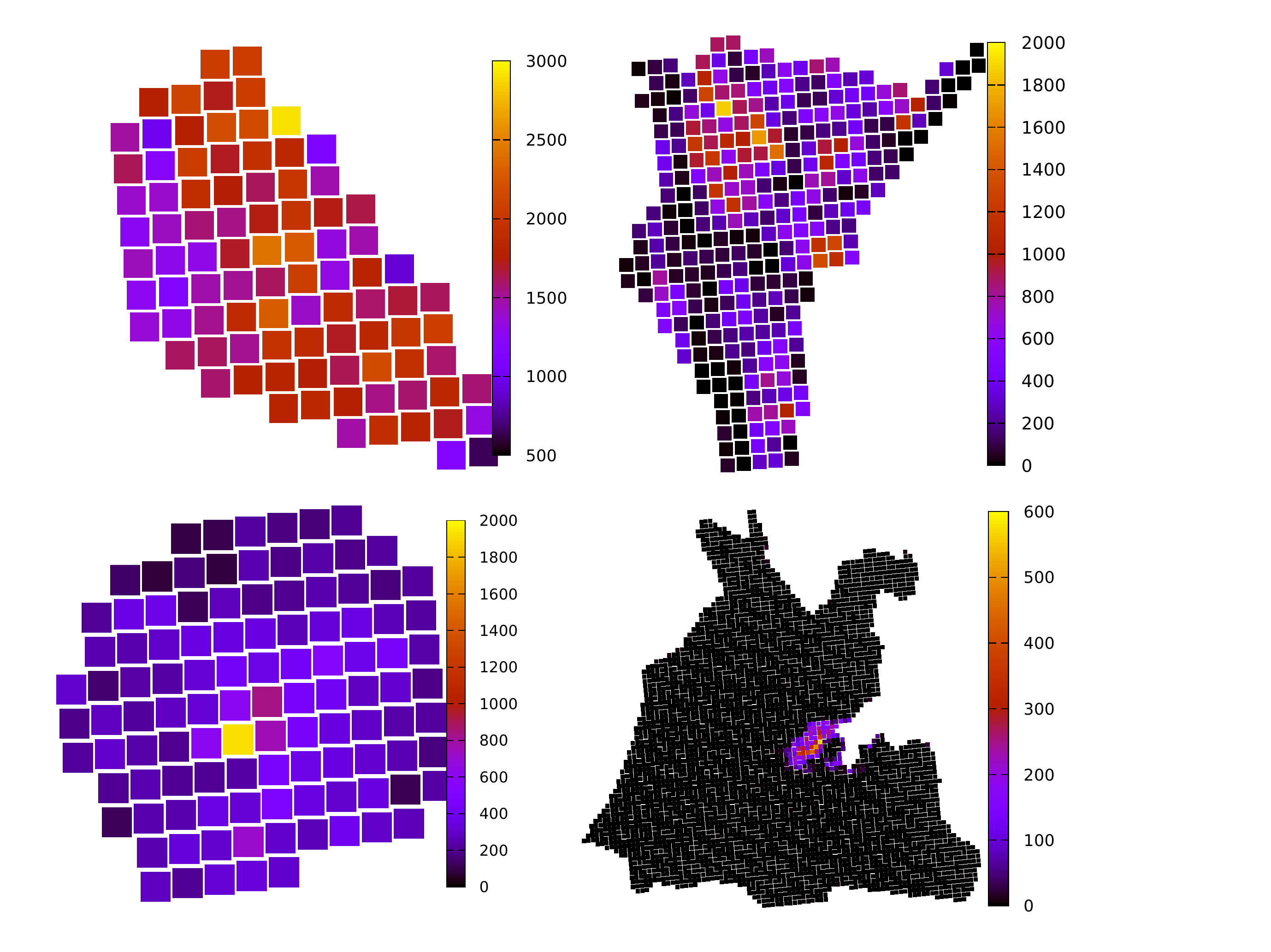}
\caption{\textbf{Representative cities for each of the four clusters.} We show here four cities, all belonging to the urban area of Paris, and which represent the characteristics of the four clusters. (top left) The 11th arrondisment of Paris, a representative of the blue cluster, of cities homogeneous and spread (or delocalized). (top right) The city of Saint-Denis, a representative of cities heterogeneous and spread (red cluster), and which are typically polycentric. (bottom left) The city of Houilles, a representative of the black cluster which contains heterogeneous and compact cities, typically monocentric cities. (bottom right) The city of Fontainebleau, a representative of the green cluster which contains very heterogenous and compact cities (typically monocentric cities too). }
\label{fig:fig13}
\end{figure}

In Fig.~\ref{fig:fig13} we show representative cities of each individual cluster, which are all in the Paris urban area. As a representative of the first (blue) cluster, we show in the top left corner of the figure the  11th district (arrondissement) of Paris. The population density is very high in all grid cells, resulting in a quite homogeneous distribution. The high populated areas are spread all over the surface of the municipality, resulting in a high value for the spreading index. The city of Paris, is another representative of this cluster. A representative element of the second (green) cluster, is shown in the bottom right of Fig.~\ref{fig:fig13} and is the municipality of Fontainebleau. The majority of the surface of the municipality is covered by a forest, and the populated area is extremely concentrated in a small part of the municipality. As a result, the population distribution is extremely heterogeneous and localized in a small region. The city of Marseille is another representative of this cluster. For the two intermediate cluster, we have in the top right of Fig.~\ref{fig:fig13} the city of Saint-Denis, a representative of the red cluster. In Saint-Denis, the population distribution is quite heterogeneous, but is particularly delocalized, having the population hotspots spread around the surface of the municipality. We can notice how this two features, the heterogeneity and the spreadness, result in a population spatial distribution that deviates significantly from monocentricity. On the contrary, the small municipality of Houilles, in the bottom left corner of Fig.~\ref{fig:fig13} is a representative of the black cluster. Here, the population distribution has a medium Gini coefficient and the population hotspots are clearly localized in the center of the city. Cities in this black cluster, which are more compact than one would expect from their heterogeneity level are typically monocentric. The cities of Lyon and Bordeaux are two other representatives of this black cluster. 

\section{Discussion}

We provided evidences that the minimal characterization of the distribution of local population densities in cities can be described along two dimensions: the heterogeneity of the distribution and the spatial location of highly populated areas. 
First, fluctuations of the local density can be very important and we should distinguish almost homogeneous cities from highly heterogeneous ones. This is easily characterized by an indicator such as the Gini coefficient $G$, or equivalently by the dispersion or the entropy. The second important dimension is the spatial organization of the population. This is essentially relevant for high heterogenous density distribution, and we propose a dispersion index $\eta$ that characterizes the degree of localization of highly populated areas, the hotspots. We used a non-parametric, distribution dependent method to identify the population hotspots. As far as population density is concerned, we argue that these two dimensions are enough to characterize the organization of a city and allows to represent different types of cities.  We discuss the relevance of this approach on approximately $4,500$ municipalities belonging the the $10$ largest urban areas in France, for which we have high resolution data. Representing these cities in the plane $(G,\eta)$ allows us to construct families of cities. Focusing on cities with population larger than $10,000$ inhabitants, we find that the more heterogeneous cities are and the more compact they appear. Secondly, we find that we can classify cities in four large categories: (i) first, homogeneous and dispersed cities where the density fluctuations are small, (ii) very heterogeneous cities with a compact organization of large densities areas. The last two groups comprise heterogeneous cities with (iii) a monocentric organization or (iv) a more delocalized, polycentric structure. We believe that integrating these two parameters in econometric analysis could improve our understanding of the impact of urban form on various socio-economical aspects.

{\bf Acknowledgements}

We thank A. Pavard, A. Bretagnolle for preparing the data and for many discussions. We also thank A. Bres, M. Breuill\'e, J. Le Gallo, R. Le Goix, C. Grivault for stimulating discussions. We thank the Soci\'et\'e du Grand Paris (SGP) for financial support.

\end{document}